\documentclass[%
 reprint,
 amsmath,amssymb,
 aps,
]{revtex4-2}

\usepackage{graphicx}
\usepackage{dcolumn}
\usepackage{bm}

\begin{document}

\preprint{APS/123-QED}

\title{Large optical forces on 
a 
barium monofluoride 
molecule using laser pulses
for stimulated absorption 
and emission: 
A full
density-matrix
simulation}
\author{A. Marsman}
\author{D. Heinrich}
\author{M. Horbatsch}
\author{E. A. Hessels}
 \email{hessels@yorku.ca}

\affiliation{%
 Department of Physics and Astronomy,
 York University
}%

\collaboration{EDM$^3$
Collaboration}

\date{\today}

\begin{abstract}
A full 
density-matrix
simulation is performed
for 
optical deflection of
a 
barium monofluoride
(BaF)
beam.
Pairs of 
counter-propagating
laser pulses are used for 
stimulated absorption
followed by 
stimulated emission.
The scheme 
produces a force 
which is nearly
an order of magnitude 
larger than
that obtainable
using
continuous-wave
laser deflection,
and yields a
force-to-spontaneous-decay
ratio which is more
than an order of magnitude
larger.
The large reduction in 
spontaneous decay is 
key to optical deflection
of molecules,
where branching ratios
to other  
vibrational 
states 
do not allow for
cycling transitions.
This work is part of an effort
by the 
EDM$^3$
collaboration to 
measure the 
electric dipole moment 
of the electron
using 
BaF 
molecules
embedded in a 
cryogenic argon solid.
Deflection of 
BaF
molecules will  
separate them
from the other
ablation products
coming from a 
buffer-gas-cooled
ablation source, 
before embedding them
into the 
argon solid.
Our simulations show that 
sufficiently large 
deflections for this 
separation are 
feasible.
\end{abstract}

\maketitle


\section{\label{sec:intro}Introduction}

Optical forces on atoms
produced 
by laser beams 
have 
proven to 
be an essential tool
in atomic physics 
since they were first
suggested 
\cite{ashkin1970acceleration}
and 
implemented
\cite{schieder1972atomic}.
Most often, 
these forces 
result from the absorption of 
one quantum 
($\hbar \vec{k}$)
of momentum,
followed by a spontaneous decay.
This process can lead
to a force as large as
${F}_0=\hbar {k}/(2 \tau)$,
corresponding to
a change of
one quantum of 
momentum every two lifetimes.

Alternative atomic laser force
schemes
use stimulated emission, 
in which 
$\hbar \vec{k}$
of momentum is imparted to 
the atom upon excitation, 
with a second 
$\hbar \vec{k}$
added to the 
atom's
momentum by an 
oppositely-directed
laser beam through
stimulated emission back 
down to the ground state.
Such 
stimulated-emission 
schemes allow for  
forces larger than
${F}_0$,
but at the expense of 
having to use
more complicated setups
involving multiple 
laser beams and 
time-varying 
optical fields.
One important example of a 
stimulated-emission 
force on atoms is the
bichromatic 
force
first demonstrated 
by Grimm, 
et al.
\cite{grimmObservation1990}
and further developed 
by others
\cite{sodingShort1997,williamsMeasurement1999,WilliamsBichromatic2000,PartlowBichromatic2004,ChiedaBichromatic2012}.
The 
bichromatic
force uses 
two 
pairs of 
counter-propagating
laser beams that are
offset from each other in 
frequency.
The frequency offset 
leads to beat notes
in both directions,
and,
at the location of an 
atom,
this beating leads to 
pulses 
(offset in time) 
arriving
from each direction.

Another, 
more straightforward
scheme for
a
stimulated-emission 
optical force
uses 
pairs of pulses
from 
counter-propagating 
laser beams
to drive stimulated
absorption and emission.
This scheme has been implemented 
for atoms by 
Long, et al.
\cite{long2019suppressed}.
In the present work,
we consider
optical forces on 
molecules 
and
simulate
a 
stimulated-emission
scheme that uses
laser pulses.
We specialize to the  
barium monofluoride
(BaF)
molecule,
the molecule 
being used by the 
EDM$^3$
collaboration to 
make an 
ultraprecise 
measurement
\cite{vutha2018orientation}
of the electric
dipole moment of the electron.
For the measurement,
BaF 
molecules 
produced by 
a 
buffer-gas-cooled
laser-ablation 
source
are embedded
in
solid 
argon. 
It is necessary to 
separate 
the
BaF 
molecules 
from the
other ablation products 
via a deflection
in order to produce
an uncontaminated solid.
For our geometry,
a deflection of 
nearly
3~m/s
(which corresponds to 
1000
quanta of photon momentum)
is required. 
This deflection is much 
larger 
than
demonstrated 
\cite{kozyryev2018coherent,galica2018deflection}
deflections of 
other molecules 
using the
bichromatic 
force.

Molecules present  
major challenges for 
optical forces
\cite{fitch2021laser}.
The first difficulty results from  
the presence of 
the 
vibrational 
levels. 
Each time the molecule
undergoes a spontaneous 
emission, 
it has a branching ratio
for returning to a 
$v>0$
level of the ground 
electronic state.
For some molecules
(including BaF)
there are also 
branching ratios
to intermediate electronic 
states.
A set of 
repump 
lasers is necessary to 
return the molecules to
the 
appropriate 
state if cooling 
is to continue.

\begin{figure}
\centering
\includegraphics[width=\linewidth]{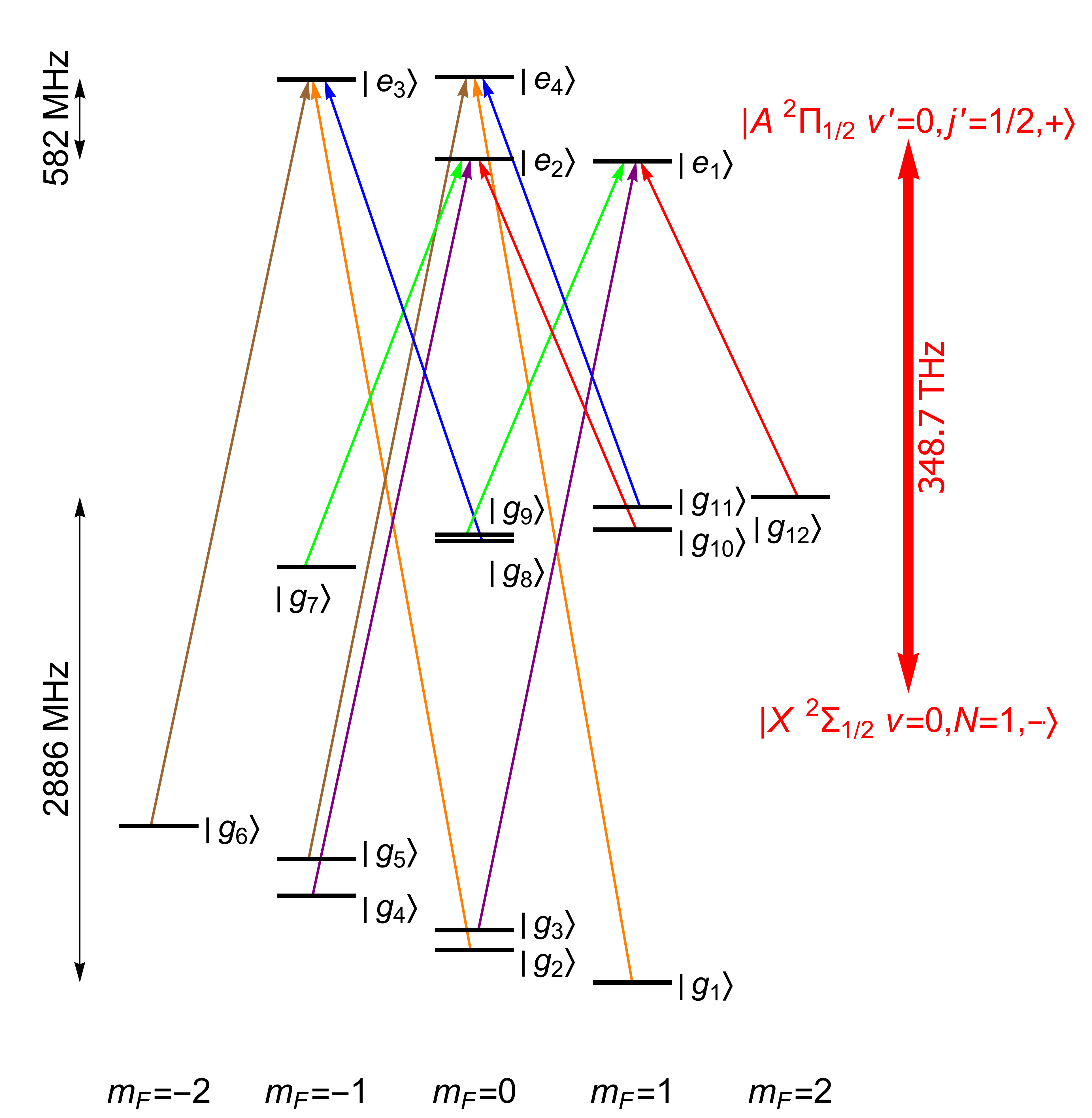}
\caption{
\label{fig:energyLevelsv0} 
(color online)
The BaF energy levels 
for the 
$X\,^2\Sigma_{1/2}\ v$$=$$0,\ N$$=$$1$
negative parity state
and
the
$A\,^2\Pi_{1/2}\ v'$$=$$0, j'$$=$$1/2$
positive parity state,
with a  
1000-gauss
magnetic field.
The magnetic field
resolves individual
transitions between 
substates, 
and
the arrows
show the twelve
transitions being driven.
Due to pairs of
similar laser frequencies,
these transitions are driven
with six 
laser
frequencies,  
as indicated by the six
colors of the arrows
and as detailed in 
Table~\ref{tab:twelveTransition}.}
\end{figure}

Additionally, 
for the molecules
most suitable for cooling
(including BaF),
the required transition has
a larger number 
$n_{\rm g}$
of 
substates
in the ground state compared 
to the number 
($n_{\rm e}$)
in the
excited state. 
This situation is illustrated 
in 
Fig.~\ref{fig:energyLevelsv0},
where the 
BaF
cooling-transition
states
\cite{hao2019high}
(the
$X\,^2\Sigma_{1/2}\ v$$=$$0, N$$=$$1$
negative parity state
and the
$A\,^2\Pi_{1/2}\ v'$$=$$0, j'$$=$$1/2$
positive parity state)
are shown.
Including the 
hyperfine 
structure, 
the ground state consists
of 
$n_{\rm g}=12$
substates,
whereas the excited state consists of 
$n_{\rm e}=4$
substates.
Having 
$n_{\rm g}$$>$$n_{\rm e}$
has two adverse effects on 
laser cooling.

The first effect
is that a laser of fixed
polarization will quickly
pump the population into 
a dark state
(a linear combination 
of the
$n_{\rm g}$
states that cannot be 
excited by this 
polarization), 
thereby shutting off
the optical force.
These dark states 
can be 
destabilized
\cite{berkeland2002destabilization}
by,
for example,
using a
frequency-modulated
laser
\cite{kaebert2021characterizing}
or by using an
appropriately-oriented
magnetic field 
\cite{shuman2010laser}
which causes 
the molecules 
to 
Larmor 
precess
into 
bright states.
Typically,
field
strengths on 
the order 
of 
10~gauss
are used for 
this 
precession.
Our approach for overcoming
the 
dark-state 
problem
is to use a much larger
magnetic field 
(of the order of 
1000~gauss)
that 
resolves the individual 
substates
(as shown in 
Fig.~\ref{fig:energyLevelsv0}) 
and then to use laser pulses to
separately address 
individual
substates.

The 
second effect
of having 
$n_{\rm g}$$>$$n_{\rm e}$
is that 
the force is smaller
even when the 
dark-state 
problem is addressed.
If, 
for example, 
there is no 
dark-state
accumulation, 
and the laser light 
is effective at equalizing the 
$n_{\rm g}+n_{\rm e}$ 
state populations,
the  
maximum force
possible 
for a spontaneous-decay force
is 
$
{F}_{\rm m}
=
2 {F}_0 n_{\rm e}/(n_{\rm e}+n_{\rm g})
$.
A similar decrease is 
also found for 
stimulated-emission 
forces.

\section{Optical deflection scheme
\label{sec:scheme}}

\begin{figure}
\centering
\includegraphics[width=3.5in]{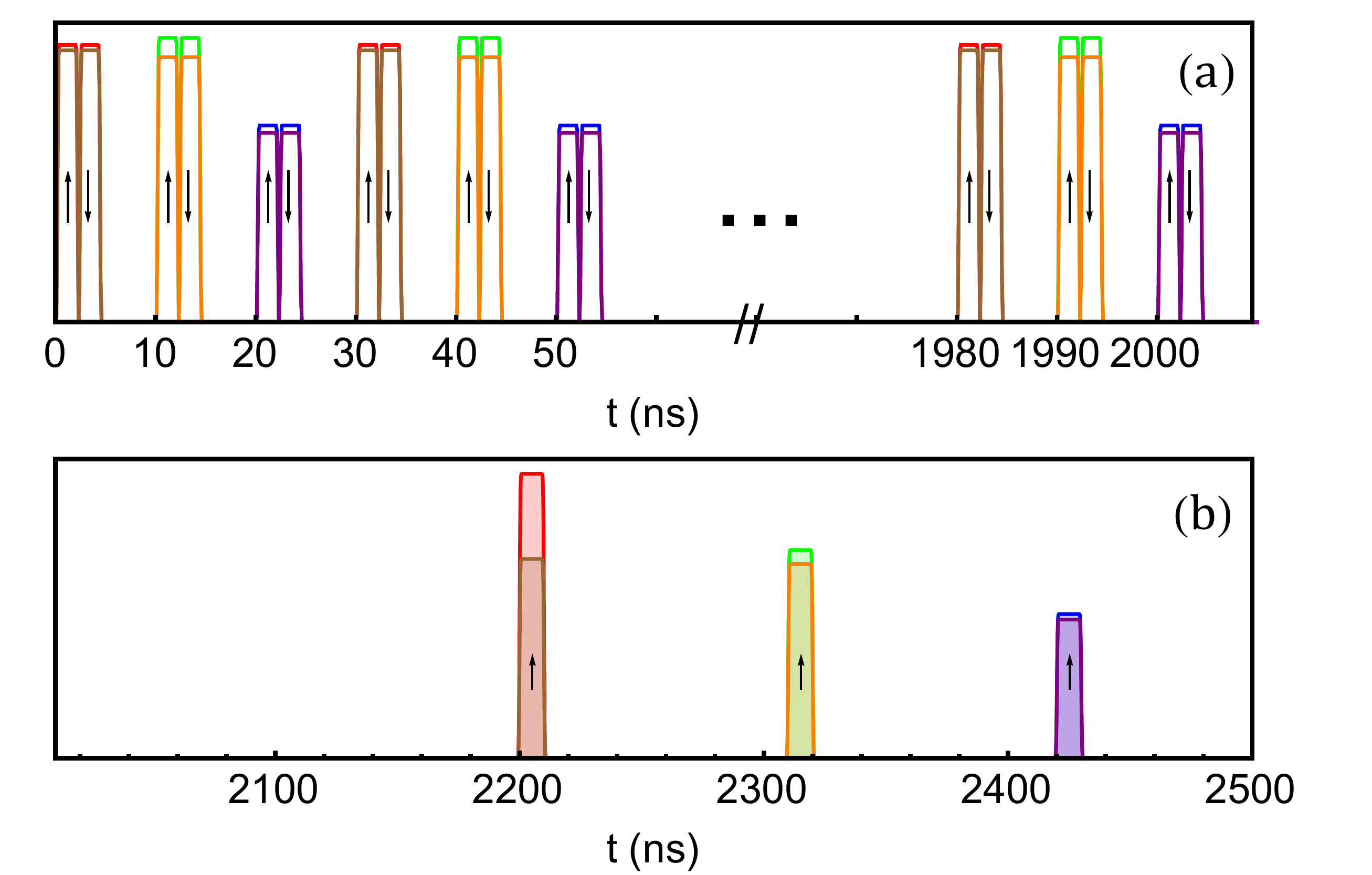}
\caption{
\label{fig:timing2us} 
(color online)
The timing diagram 
for the 
laser pulses.
Panel~(a)
shows 
the first 
2~$\mu$s
of deflection.
A 
2-ns
$\pi$
pulse from the 
upward-directed 
laser beam is immediately
followed by a similar 
$\pi$
pulse from the 
downward-pointing laser 
beam. 
Two frequencies 
are present in each 
pulse, 
as indicated by the 
colors here and 
the matching colors 
in 
Fig.~\ref{fig:energyLevelsv0}.
These two frequencies 
address populations in 
four of the 
twelve 
$|g_i\rangle$
states. 
The two pulses are 
followed 
10~ns
and 
20~ns
later by  
similar pulses 
that address 
the other 
$|g_i\rangle$
states, 
with this 
30-ns
sequence being 
repeated 
67 
times.
After these 
pulses,  
single 
10.5-ns
pulses for each of the 
analogous laser transitions
from the 
$v=1$
state
serve
to
repump
the population back into 
the 
$v=0$,
as shown in 
panel~(b).
The sequence shown in both panels 
is repeated 
approximately 
15 times
during the time 
the molecule passes
through the laser
beams.
} 
\end{figure}

Our proposed
scheme for optical deflection
uses a series of 
laser pulses,
each with a 
duration of 
2~ns,
as shown in 
Fig.~\ref{fig:timing2us}(a).
A first pulse,
from an 
upward-directed 
laser beam 
is immediately
followed by a second pulse
from a 
downward-directed 
beam.
The intention is
for the first pulse 
to move 
ground-state
population
up to the excited 
state, 
and for the second pulse
to stimulate this 
population back down
to the ground state.
Each set of two pulses
in 
Fig.~\ref{fig:timing2us}(a)
includes two separate 
frequencies
(as indicated by the 
two 
colors 
of pulses),
and these two frequencies
drive four transitions,
as indicated by the four
arrows with these two 
colors
in 
Fig.~\ref{fig:energyLevelsv0}.
A 
1000-gauss
magnetic field
(along the axis of the laser beams)
lifts the degeneracies between
$|g_i\rangle$
states,
allowing these four 
isolated
transitions
to be driven, 
with very little
accidental transfer due to 
other 
$|g_i\rangle$-to-$|e_j\rangle$
transitions.
The intensities at 
each frequency are 
chosen to meet the 
$\pi$-pulse
condition, 
so the population of 
four of the 
$|g_i\rangle$
states of 
Fig.~\ref{fig:energyLevelsv0}
are efficiently excited to the 
four  states,
$|e_1\rangle$,
$|e_2\rangle$,
$|e_3\rangle$
and
$|e_4\rangle$,
and then returned
to the original
four
$|g_i\rangle$
states.

If the ground state
consisted of only 
four substates, 
this pair of pulses 
could be repeated 
and would lead
to two quanta 
of momentum being 
imparted
during each cycle.
With twelve substates,
however,
spontaneous decay
would 
move population
out of these four states
into the other
eight 
$|g_i\rangle$
states,
and these eight states
would be dark states.
The 
dark-state
problem
can be overcome by having the 
next pair of pulses 
(those shown in purple 
and 
blue
in 
Fig.~\ref{fig:timing2us}(a))
address
four other 
$|g_i\rangle$
states
using two frequencies that 
drive the four 
purple and blue
transitions shown in 
Fig~\ref{fig:energyLevelsv0}.
A third pair of pulses then
addresses the remaining
four 
$|g_i\rangle$
states
(orange and brown in 
Figs.~\ref{fig:energyLevelsv0}
and
\ref{fig:timing2us}(a)).

If there were no
spontaneous decay,
this cycle of pulses 
(the first 
30~ns
of 
Fig.~\ref{fig:timing2us}(a))
could
be repeated indefinitely
without any concern about 
dark states.
If, 
however,
a spontaneous decay
occurs,
for example,
at 
2~ns
in 
Fig.~\ref{fig:timing2us}(a),
the second pulse, 
which is intended to
stimulate the population
back down to a 
$|g_i\rangle$
state would instead
excite it up to 
an 
$|e_j\rangle$
state,
thus imparting 
a quantum of momentum
in the opposite direction.
Furthermore, 
all of the following pulses 
would similarly perform
the reverse role and 
also impart the 
wrong sign of momentum 
to the molecule.
In this scenario,
the molecule continues
to experience this 
opposite force until another 
spontaneous decay
puts the scheme back on 
track.
The delay time between  
pulse pairs
(e.g., between the pulse 
centered
at
3~ns
and the pulse 
centered at 
11~ns
in
Fig.~\ref{fig:timing2us}(a))
makes the spontaneous 
decays that correct the 
scheme back to the intended
cycle happen 
at a faster rate
than spontaneous decays 
that reverse the force,
which can only happen if 
the spontaneous decay
occurs between the 
closely-spaced 
pulses
(e.g., the pulses
centered at 
11 
and 
13~ns
in 
Fig.~\ref{fig:timing2us}(a)).

The net effect is that 
most often
(about 
80\%
of the time, 
which comes 
from the 
8-ns
center-to-center
pulse separation
for the delay
compared to the
10-ns 
cycle between
pulse pairs)
the force is in the 
correct direction
and 
20\% of the time 
it is in the 
wrong direction.
This leads to a net
force corresponding to 
0.6
(80\%
minus
20\%)
times
two quanta
of momentum
divided by three
per 
10~ns.
The division by 
three results from 
the fact that only 
one third of the population
is addressed during each
10-ns
period.
The net force predicted 
is nine times larger than
the force 
${F}_m$
possible for 
spontaneous-decay
forces 
(which could only be 
achieved if the 
dark-state problem
were 
eliminated).
Compressing the time scale 
of 
Fig.~\ref{fig:timing2us}
would lead to even larger
forces, 
but the shorter pulse duration
may require a larger magnetic
field to sufficiently resolve
the energy levels of 
Fig.~\ref{fig:energyLevelsv0}
given the broader spectrum
inherent in the shortened pulse
duration.
Higher-intensity
laser light would 
also
be needed to 
satisfy the
$\pi$-pulse
criterion.

As will be seen in the following 
section, 
most of the population resides
in the ground states at all times.
This reduces the spontaneous decays
and therefore reduces the population
lost to 
$v=1$
and
other states.
As a result, 
only one 
repump 
laser is required 
in the current scheme.

\section{Complete density-matrix simulation \label{sec:densityMatrix}}

\begin{figure}
\centering
\includegraphics[width=2.5in]{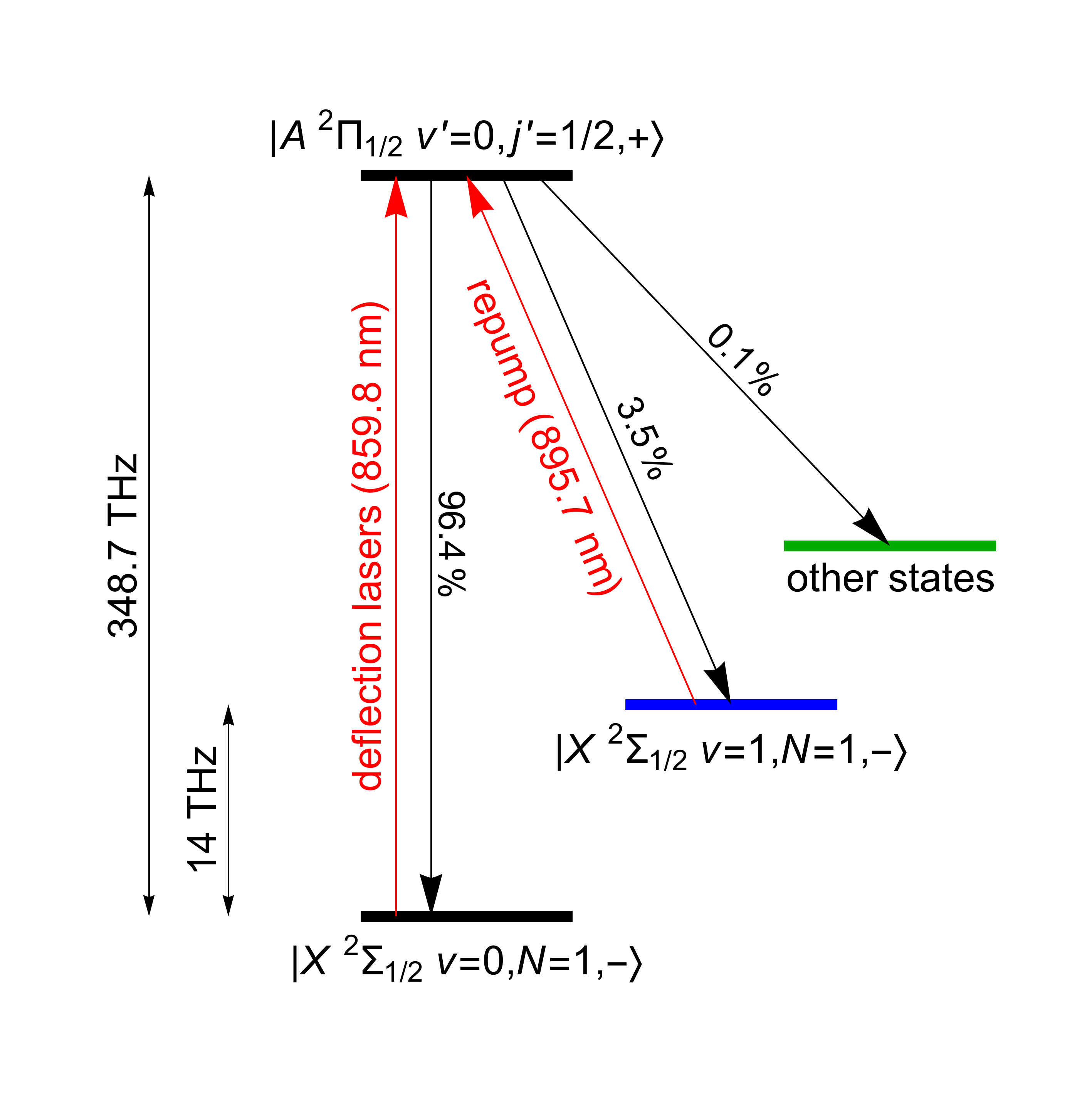}
\caption{
\label{fig:branching} 
(color online)
The branching ratios
for the 
$A^2\Pi_{1/2}$
state
and laser wavelengths
needed for our 
cooling scheme.
} 
\end{figure}

In this section we present a 
density-matrix
simulation of the 
deflection scheme.
Since a
BaF 
molecule in the 
$A\,^2\Pi_{1/2}(v=0)$
state has
a branching ratio
\cite{hao2019high}
of
96.4\%
for decay to the 
$X\,^2\Sigma_{1/2}(v=0)$
state
(see 
Fig.~\ref{fig:branching}),
the 
sixteen states
of 
Fig.~\ref{fig:energyLevelsv0}
form an almost closed system.
However, the 
3.5\%
branching ratio
\cite{hao2019high}
to
the 
$X\,^2\Sigma_{1/2}(v=1)$
state
requires 
an additional twelve
ground states 
($|g_{13}\rangle$ 
through
$|g_{24}\rangle$
that are the 
$v=1$
analogues 
to 
$|g_{1}\rangle$ 
through
$|g_{12}\rangle$
of 
Fig.~\ref{fig:energyLevelsv0})
to be 
included in the simulation.
A 
repump 
laser
is required to avoid
accumulation of population 
in these twelve states.
The remaining branching 
ratio of 
0.1\%
\cite{hao2019high}
includes
spontaneous decay to 
the
$X\,^2\Sigma_{1/2}(v$$>$$1)$
states 
and the 
$A'\,^2\Delta_{3/2}$
state.
Spontaneous decay to 
the latter state
causes the molecule to 
change its parity upon
its second decay down 
to the 
$X\,^2\Sigma_{1/2}$
state.
As will be shown,
no
repump
lasers are needed for these states,
thus allowing them to be treated
as a single additional state
($|g_{25}\rangle$),
which acts as a 
dark state 
in which population 
accumulates over time. 

In all, 
29
states are included in the 
simulation:
25 
ground states and 
4
excited states.
The 
time evolution of
the 
density matrix 
elements can be determined 
by numerically
integrating their
differential equations
during the time period 
when the molecule
passes through the 
laser beams:

\begin{eqnarray}
\frac{d\rho_{g g'}}{dt}&=&
i \omega_{g'\!g} \rho_{g g'}
+ \frac{i}{2}
\sum_e 
(\Omega_{e g'} \rho_{g e}
-\Omega_{g e}  \rho_{e g'}) 
\nonumber
\\
&&+\sum_{e,e'} \gamma_{e g e'\!g'} \rho_{e e'};
\nonumber
\\
\frac{d\rho_{e e'}}{dt}&=&
i \omega_{e'\! e} \rho_{e e'}
+\frac{i}{2} \sum_{g} 
(\Omega_{g e'}\rho_{e g}
-\Omega_{e g} \rho_{g e'})
\nonumber
\\
&&-\frac{1}{2} \sum_{g,e''} 
(\gamma_{e g e''\! g} \rho_{e''\! e'}
+\gamma_{e''\! g e'\! g} \rho_{e e''});
\nonumber
\\
\frac{d\rho_{g e}}{dt}&=&
i(\omega_{e g}-\omega_v) \rho_{g e}
+\frac{i}{2}\sum_{g'} \Omega_{g'\!e} \rho_{g g'}
\nonumber
\\
&&-\frac{i}{2}\sum_{e'} \Omega_{g e' } \rho_{e'\!e}
-\frac{1}{2}\sum_{g',e'} \gamma_{e'\!g'\!eg'} \rho_{g e'},
\label{eq:denMatrix}
\end{eqnarray}
where the indices 
$g$
and 
$e$
represent the 
25
ground 
and 
4
excited states,
respectively.
Here,
$\hbar \omega_{ab}=E_a-E_b$
is the energy 
difference between 
two states
$|a\rangle$
and 
$|b\rangle$,
in the 
1000-gauss
magnetic field.
These energies are calculated
using the methods 
detailed in the 
appendix of 
Ref.~\cite{kaebert2021characterizing}
and are listed in 
Table~\ref{tab:energies}
in the appendix 
of this paper.
The difference
between the 
average energy of 
the 
$|e\rangle$
states
and the 
$|g\rangle$
states for 
each value of
$v$
is denoted by 
$\omega_{v}$
($v=0,1$).
The 
rotating-wave 
approximation is 
used (twice) 
to avoid the 
fast oscillations 
at the optical 
frequencies 
$\omega_0$
and
$\omega_1$.

We use the complete
formulation for spontaneous
decay
\cite{cardimona1983spontaneous,marsman2012shifts}
which includes 
quantum-mechanical 
interference 
from the decay process
using  
\begin{equation}
\gamma_{e g e'\!g'}
=
\frac{\omega^3}{3 \pi \epsilon_0 \hbar c^3}
\vec{d}_{ge} \cdot \vec{d}_{e'\!g'}.
\end{equation}
The diagonal 
elements 
$\gamma_{e g e g}$
are equal to the 
branching ratio 
times 
$1/\tau$.
The dipole matrix elements
can then be deduced from 
the measured lifetime of the 
$A\,^2\Pi_{1/2}$
state  
($\tau=$
57.1~ns
\cite{aggarwal2019lifetime}),
along with the 
branching ratio 
\cite{hao2019high} 
to each vibrational state 
and ratios 
for transitions
from an individual 
Zeeman 
state
$|e_j\rangle$
to
$|g_i\rangle$.
These latter ratios are 
calculated using the methods
described in 
Ref.~\cite{kaebert2021characterizing}.
The values of the
electric dipole matrix elements
used are listed 
in 
Table~\ref{tab:matrixEls} in 
the appendix.

The 
Rabi
frequencies
in 
Eq.~(\ref{eq:denMatrix})
are given by
\begin{eqnarray}
\label{eq:fRabi}
\Omega_{g e}(t)
\!&=&\!
\vec{d}_{ge} \cdot
\sum_p
\frac{\hat{\epsilon}_p E_{0p}(\vec{r},t)}{\hbar}
e^{i [(\omega_p-\omega_v)t - \vec{k}_p \cdot \vec{r} + \phi_p ]}
\Bigg\rvert_{\vec{r}=\vec{r}_m(t)}.
\nonumber
\\
&& 
\end{eqnarray}
Here,
$\vec{d}_{ge}$
is the 
electric dipole matrix element
between states 
$g$
and
$e$
and 
the sum is 
over all laser 
fields,
which includes
the upward and 
downward beams
of 
Fig.~\ref{fig:timing2us}(a),
with all six
frequency components
(as detailed in 
Table~\ref{tab:twelveTransition}),
as well as six 
repump
frequencies,
as indicated in 
Fig.~\ref{fig:timing2us}(b).
The frequencies,
wavevectors,
phases,
polarizations
and 
amplitudes 
are represented by 
$\omega_p/(2 \pi)$,
$\vec{k}_p$,
$\phi_p$,
$\hat{\epsilon}_p$
and
$E_{0p}(\vec{r},t)$,
respectively.
The latter includes
both the 
time profile 
of the laser pulses
(Fig.~\ref{fig:timing2us})
and the spatial profile
of the laser beam, 
which is taken to be 
a 
top-hat 
shape 
(as, 
e.g.,
in 
Ref.~\cite{ma2011improvement})
approximated by an 
elliptical 
super-gaussian
function
of the form
\begin{equation}
e^{(-x^2/w^2-y^2/h^2)^5},
\end{equation}
where the values of  
$w$
and 
$h$
are chosen such that the 
full width at half maximum 
of the beam 
is 
5~mm 
in width
and 
1~mm 
in height.
The values of 
$E_{0}$
used for each of the beams
is given in 
Table~\ref{tab:twelveTransition},
and these correspond to 
peak laser powers
(during the 
2-ns
pulses)
of approximately
2~W.

\begin{table}[]
\begin{ruledtabular}
\centering
\begin{tabular}{lcccc}
transitions&
polar-&
Figs.&
$\Delta f$&
$E_0$
\\
&
ization&
\ref{fig:energyLevelsv0}\&\ref{fig:timing2us}(a)&
(MHz)&
(V/cm)\\
\hline
$|g_{12}\rangle\to|e_1\rangle$, $|g_{10}\rangle\to|e_2\rangle$&
$\sigma^-$&
red&
$-1696$&
208\\
$|g_6\rangle\to|e_3\rangle$, $|g_5\rangle\to|e_4\rangle$&
$\sigma^+$&
brown&
$+1615$&
208\\
$|g_9\rangle\to|e_1\rangle$,$|g_7\rangle\to|e_2\rangle$&
$\sigma^+$&
green&
$-1659$&
217\\
$|g_2\rangle\to|e_3\rangle$, $|g_1\rangle\to|e_4\rangle$&
$\sigma^-$&
orange&
$+1739$& 
203\\
$|g_{11}\rangle\to|e_4\rangle$, $|g_8\rangle\to|e_3\rangle$&
$\sigma^-$&
blue&
$-1104$&
150\\
$|g_4\rangle\to|e_2\rangle$, $|g_3\rangle\to|e_1\rangle$&
$\sigma^+$&
purple&
$+1104$&
145\\
\end{tabular}
\end{ruledtabular}
\caption{The parameters 
used for the twelve
laser transitions of 
Fig.~\ref{fig:energyLevelsv0}.
}
\label{tab:twelveTransition}
\end{table}

Both
the 
$\vec{r}$
dependence 
in this profile 
and the 
$\vec{r}$
in the complex
exponential of 
Eq.~(\ref{eq:fRabi})
are evaluated at the 
molecular position
$\vec{r}_m(t)$.
The 
trajectory 
of the molecule
is obtained from 
its original position
(assumed to be
3.25~mm
before the 
axis of the laser beams),
its initial velocity
(assumed to be 
150~m/s
--
a typical speed
for a
BaF 
molecule
in a
4-kelvin
helium-buffer-gas-cooled
beam
\cite{truppe2018buffer}
--
directed towards the
axis of the laser beams)
and the
force
obtained
\cite{cook1979atomic}
from
Ehrenfest's
theorem:

\begin{equation}
\vec{F}(t)=-  \hbar \sum_{e,g} {\rm Re}
[
\rho_{e g}(t) \nabla \Omega_{g e}(\vec{r},t)
]
\Bigg\rvert_{\vec{r}=\vec{r}_m(t)}.    
\end{equation}

The density matrix 
has 
$29^2=841$
components. 
However,
we set 
$\rho_{g g'}=0$
when 
$g$
and
$g'$
do not have the same
vibrational 
quantum number
$v$,
since
the 
$i\omega_{g' g}$
term in its 
differential 
equation
causes rotations 
at an infrared 
frequency 
which causes the 
much slower 
accumulation 
in this state 
from the 
$\gamma_{e g e'\!g'}$
term to average to zero.
The remaining 
505
density matrix elements,
along with the components 
of 
$\vec{r}_m(t)$
and 
$\dot{\vec{r}}_m(t)$
are numerically integrated 
over the 
45~$\mu$s
that it takes for the 
molecules to pass 
through the 
laser profile.
These 
integrations 
are
computationally intensive
since the 
$\omega_{g'\!g}$
and 
$\omega_{e g}-\omega_v$
terms in 
Eq.~(\ref{eq:denMatrix})
cause oscillations with 
periods of less than 
1~ns
that must be 
calculated accurately.

\begin{figure}
\centering
\includegraphics[width=3.5in]{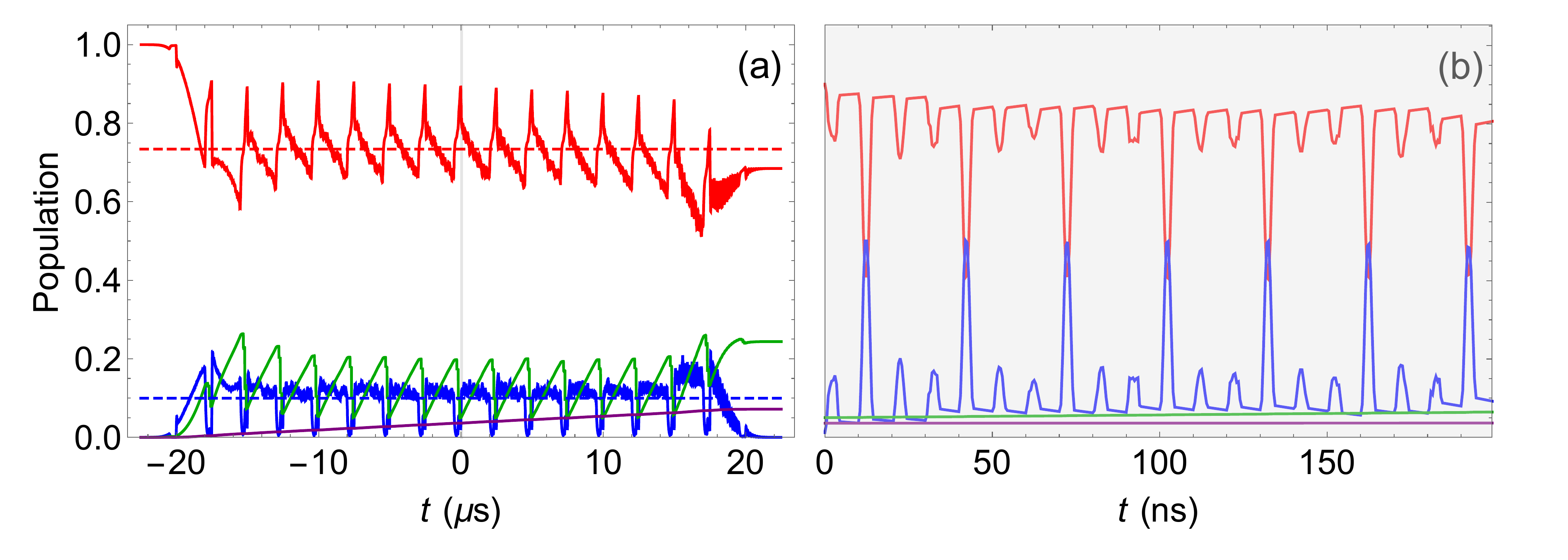}
\caption{
\label{fig:populations} 
(color online)
The total population 
of the 
$v$$=$$0$
ground states
(red),
the excited states
(blue),
the 
$v$$=$$1$ 
ground states
(green),
and all 
other states
(purple).
Panel~(a)
shows the 
time-averaged
(50-ns 
moving average)
populations 
for the entire deflection
time,
where it can be seen that
the 
$v$$=$$1$
population 
accumulates to 
18\%
over 
2~$\mu$s
of deflection pulses,
at which time 
repump 
pulses
return
the majority to the  
$v$$=$$0$ 
state.
The population 
in all other states
remains small 
throughout the 
deflection time
indicating that no 
additional
repump
lasers
are needed.
The
200-ns
segment 
shown in the narrow grey region
is expanded 
in 
panel~(b) 
without 
time averaging 
so that
the effect of the
individual 
$\pi$ 
pulses 
can be 
seen.
The larger features 
in this panel are due 
to excitations from 
$|g_1\rangle$,
$|g_2\rangle$,
$|g_7\rangle$,
and
$|g_9\rangle$,
which are 
the four states with 
the largest population
due to their 
larger branching
ratios.
} 
\end{figure}

The results of this 
integration are
given in 
Fig.~\ref{fig:populations},
where the total population
(red) 
of 
the 
twelve
$v$$=$$0$
ground states
is shown 
along with the 
total
of the 
four
excited states
(blue).
The time slice in 
panel~(b)
of the figure shows 
the individual
$\pi$-pulse
population transfers,
whereas 
panel~(a)
shows  
time-averaged
populations over the 
time taken by the 
molecule to pass through 
the laser beams.
On average,
only 
10\%
of the population
(dashed blue line)
is in an excited
state, 
which implies that,
on average,
it takes 
approximately ten
lifetimes for
each molecule to
undergo a spontaneous
emission.
This reduced emission
slows the loss of population 
to dark states;
however, 
by the end of  
each 
2~$\mu$s
of deflection laser pulses
(Fig.~\ref{fig:timing2us}(a)),
18\%
of 
the population
is in 
the 
$X\,^2\Sigma_{1/2}\ v$$=$$1, N$$=$$1$
dark states
(green in
Fig.~\ref{fig:populations}(a)).

To avoid further
population buildup
in these dark states,
six 
10.5-ns 
$\pi$
pulses
(as shown in 
Fig.~\ref{fig:timing2us}(b))
drive  
twelve transitions from each of the
twelve
$X\,^2\Sigma_{1/2}\ v$$=$$1, N$$=$$1$
states
($|g_{13}\rangle$
through
$|g_{24}\rangle$)
to one of the 
$|e_{1}\rangle$
through
$|e_4\rangle$
states.
These transitions are the exact 
$v=1$
analogues of the transitions shown
in
Fig.~\ref{fig:energyLevelsv0}
and in 
Table~\ref{tab:twelveTransition},
with the longer pulse length 
compensating for the
smaller 
electric-dipole
matrix elements
coupling these states.
The deflection and 
repump
sequence of 
Fig.~\ref{fig:timing2us}
is repeated
as the molecule 
passes through the
laser beams.

\begin{figure}
\centering
\includegraphics[width=3.5in]{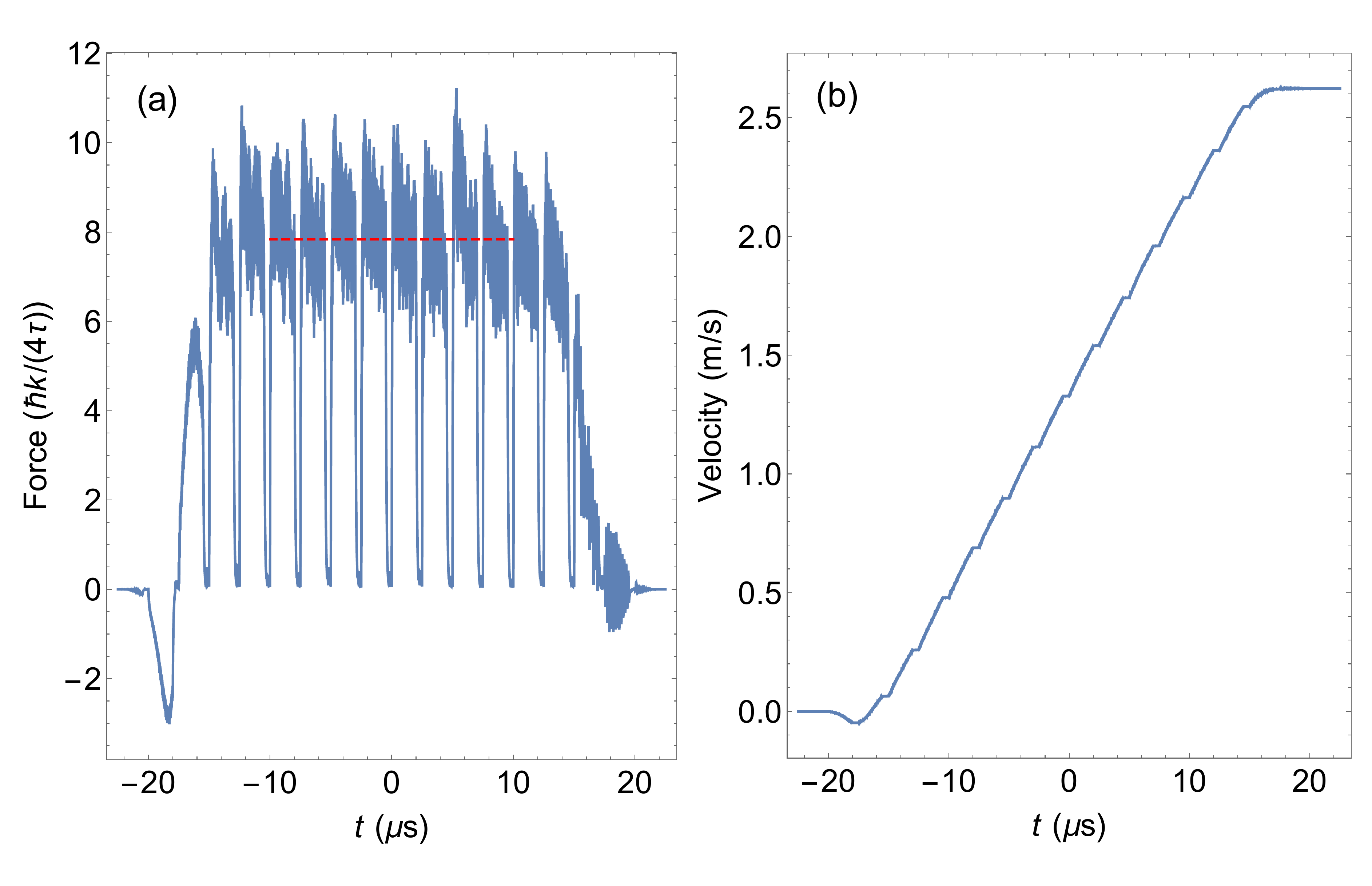}
\caption{
\label{fig:vAndF} 
(color online)
The force
(a)
given
in units
of 
$F_m$
and 
the transverse velocity 
(b)
versus time
for a 
BaF
molecule 
that is being deflected
by the scheme presented 
in this work.
} 
\end{figure}

Figure~\ref{fig:vAndF}(a)
shows the force experienced 
by the molecule as it passes
through the laser beam.
The average force 
is eight times larger 
(dashed line)
than
the  
spontaneous-emission 
force 
(${F}_m$$=$$\hbar k/(4 \tau)$),
showing that our scheme
deals effectively with dark 
states and that the stimulated
absorption and emission cycle
works almost as efficiently 
as the estimate given in 
Section~\ref{sec:scheme}.
This force is significantly larger
than demonstrated 
bichromatic 
forces on molecules
and the simulated
\cite{kogel2021laser}
bichromatic force 
on 
$^{137}$Ba$^{19}$F.

Given the potential loss to
dark states for each 
spontaneous emission,
a key figure of merit
for molecular deflection
is the number of quanta
of momentum imparted per
spontaneous emission.
The combination of the large 
force of
Fig.~\ref{fig:vAndF}(a)
and the low 
spontaneous-emission 
rate due to the 
low 
excited-state 
population of 
Fig.~\ref{fig:populations}
implies
that 
20~$\hbar \vec{k}$
of momentum 
is imparted per
spontaneous decay.
This ratio is 
20 
times larger
than that possible for 
non-stimulated 
optical deflection schemes.
Because of the 
low level of spontaneous emission, 
the effect of 
all of the other dark states
(the 
$|g_{25}\rangle$
state 
in our model)
is minimal,
with a population 
accumulation of less than
5\%,
as seen by the purple
curve in 
Fig.~\ref{fig:populations}(a).
As a result, 
the present scheme
requires only one 
repump 
wavelength.
In all, 
two laser wavelengths
are required
(859.8
and
895.7~nm),
each with 
sidebands
to provide 
six frequencies
(see 
Table~\ref{tab:twelveTransition}).

Figure~\ref{fig:vAndF}(b)
shows the transverse velocity 
of the molecule as a function
of time. 
The molecule is deflected by 
2.6~m/s
as it passes through the lasers.
This deflection is sufficient to 
fully separate 
BaF
molecules
from the other 
ablation products.
Since the lasers are 
only in resonance with the 
$^{138}$Ba$^{19}$F
isotope,
the resulting 
BaF 
beam 
will be 
isotopically 
pure.

An important practical 
consideration for any deflection
scheme is the dependence of the 
deflecting force on the laser 
parameters,
which are bound to be 
imperfectly set in 
an actual implementation 
of the deflection.
Full density matrix simulations 
are repeated multiple times
with the 
six
values of 
$E_0$ 
of
Table~\ref{tab:twelveTransition}
(and the similar six values 
for the 
$v=1$
transitions)
individually multiplied by 
random factors between
0.95 
and 
1.05.
These simulations
show that the deflection
is robust against 
variations of the 
laser field amplitude,
with the standard deviation
of the resulting deflections
($\sigma_{\rm defl}$)
being less than 
2\%.
Similarly, 
imperfect 
polarization 
for each of the 18
different laser pulses
(differing in frequency or direction)
is modelled by
replacing 
$\sigma^{\pm}$
polarization 
in 
Table~\ref{tab:twelveTransition}
with 
$\sigma^{\pm}+\epsilon\sigma^{\mp}$,
where 
$\epsilon$
is chosen randomly to be between
$-1\%$ and
$+1\%$
for each laser.
The deflection was also
robust against this variation,
with a  
standard deviation
$\sigma_{\rm defl}$
of less than
$1\%$.
Finally,
the phase of all eighteen
laser frequencies was
randomly varied
(between 
0
and 
$2\pi$)
and again the 
variation in 
deflection was small,
with $\sigma_{\rm defl}<6\%$.

\begin{figure}
\centering
\includegraphics[width=2.5in]{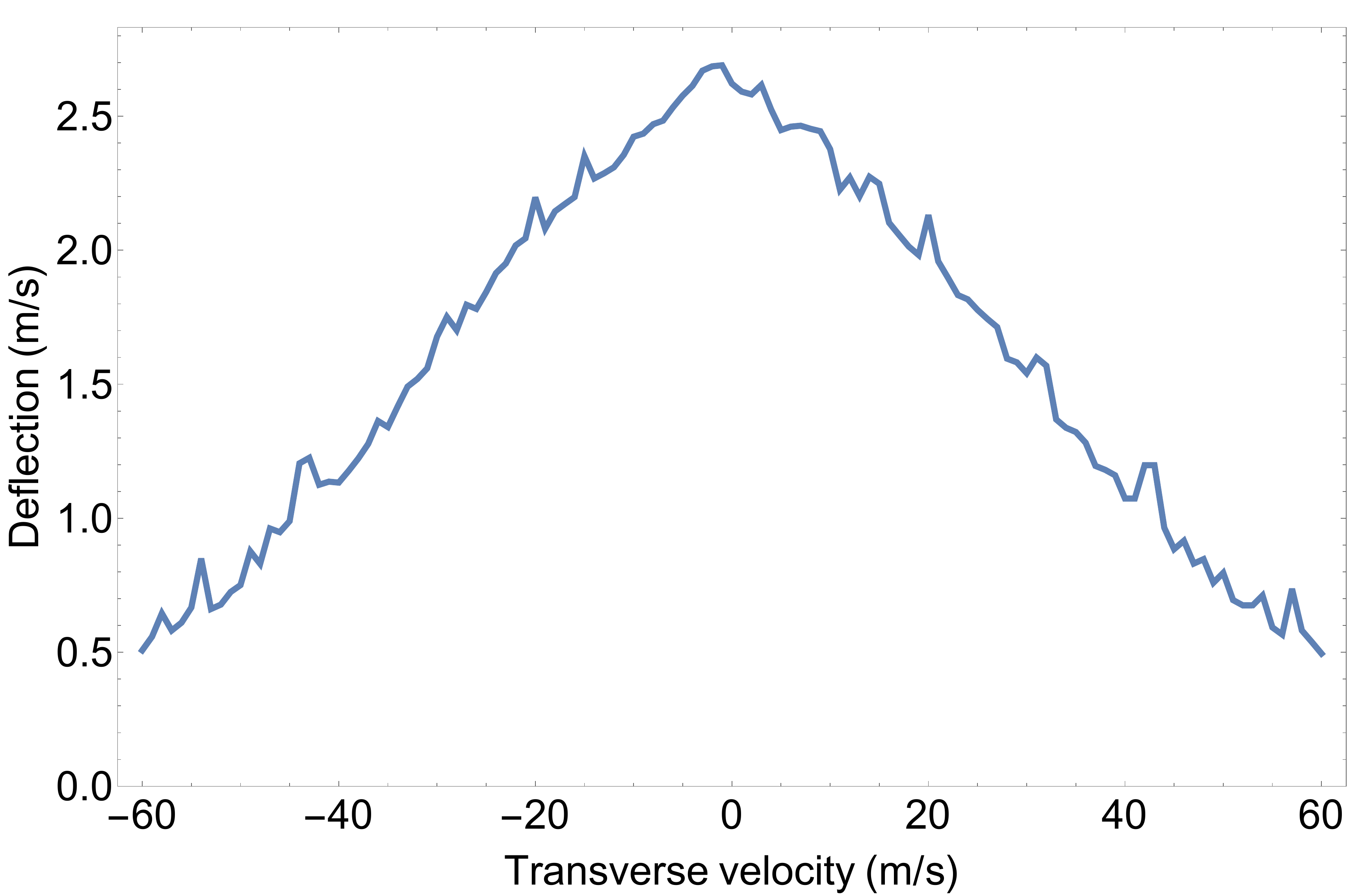}
\caption{
\label{fig:captureRange} 
(color online)
Deflection of the
BaF 
molecule versus
its initial
transverse velocity.
The scheme described in this 
work is effective over 
a large velocity range.
} 
\end{figure}

The simulations were also performed
with different initial velocities.
Changing the longitudinal component of the 
velocity along the molecular beam 
axis
has only a small effect 
on the magnitude of the 
force.
For example, 
a
7\%
change
in this longitudinal component
leads to a change in the 
force of less than 
2\%.
We also repeated the simulations
with different initial velocity 
components
transverse to the molecular
beam axis
(along the axis
of the laser beams),
while fixing the 
longitudinal component at
150~m/s.
In order to separate the 
BaF
molecules from the 
other ablation products,
we need only to communicate to
transverse velocity components
of between 
$-2$
and
$+2$~m/s.
However, 
the simulations show that
our deflection scheme
is effective over a
wide 
range 
of transverse speeds,
as shown in 
Fig.~\ref{fig:captureRange}.

\section{\label{sec:concl}
Conclusions}

In this work, 
we have performed a complete
density-matrix
simulation of 
a scheme that is capable
of providing strong 
laser deflection forces for 
molecules.
Pairs of pulses 
from 
oppositely-directed
laser beams
are used 
to stimulate population 
to an excited state 
followed by stimulated
emission.
Dark states are avoided by 
applying a magnetic field
large enough to 
resolve 
transitions between 
individual magnetic 
substates.
This scheme would allow for
a beam 
of 
$^{138}$Ba$^{19}$F
to be separated from 
other laser ablation 
products coming from 
a
buffer-gas-cooled
laser-ablation
source,
as required by the 
EDM$^3$
collaboration 
for their planned measurement 
of the electric dipole moment
of the electron 
using 
BaF
molecules embedded in an 
Ar 
solid.


\section{\label{sec:ackn}
Acknowledgements}

We acknowledge support from
the 
Gordon and Betty Moore Foundation,
the
Alfred P. Sloan Foundation,
the
John Templeton Foundation 
(through 
the 
Center for Fundamental Physics
at
Northwestern University),
the 
Natural Sciences
and Engineering Council 
of Canada, 
the 
Canada Foundation for Innovation, 
the
Ontario Research Fund
and
from
York University.
Computations for this work were 
enabled by support provided by 
the 
Digital Research Alliance of Canada,
Compute Ontario and 
SHARCNET.

\section*{Appendix}
The energies of the 
$v=0$ 
states shown in 
Fig.~\ref{fig:energyLevelsv0}
were calculated
using the methods 
detailed in 
Ref.~\cite{kaebert2021characterizing},
as were the 
electric-dipole 
matrix elements
between 
these states.
The calculated energies
are listed in 
Table~\ref{tab:energies},
and the
electric-dipole matrix 
elements are 
listed in 
Table~\ref{tab:matrixEls}.

\begin{table}[]
\begin{ruledtabular}
\centering
\begin{tabular}{cccccc}
$g$&$m_F$&$E_g$(MHz)&$e$&$m_F$&$E_g$(MHz)\\
\hline
1&1&-1464.63&1&1&-292.14\\
2&0&-1431.73&2&0&-289.66\\
3&0&-1412.34&3&-1&289.65\\
4&-1&-1377.77&4&0&292.16\\
5&-1&-1340.75\\
6&-2&-1307.81\\
7&-1&1351.63\\
8&0&1377.39\\
9&0&1384.07\\
10&1&1389.16\\
11&1&1411.43\\
12&2&1421.34\\
\end{tabular}
\end{ruledtabular}
\caption{
The relative energies 
of the ground states
$|g\rangle$
and 
excited states
$|e\rangle$
of 
Fig.~\ref{fig:energyLevelsv0}
for a 
1000-gauss
magnetic field.
}
\label{tab:energies}
\end{table}

\begin{table}[]
\begin{ruledtabular}
\centering
\begin{tabular}{ccccccccc}
\multicolumn{3}{c}{$\sigma^-$}&
\multicolumn{3}{c}{$\pi$}&
\multicolumn{3}{c}{$\sigma^+$}\\
$e$&$g$&$\langle e|d|g\rangle$&
$e$&$g$&$\langle e|d|g\rangle$&
$e$&$g$&$\langle e|d|g\rangle$\\
\hline
2&1&-0.015&1&1&-1.329&1&2&-0.001\\
4&1&0.963&2&2&-1.329&1&3&-1.350\\
3&2&0.964&4&2&-0.005&2&4&1.350\\
3&3&-0.016&2&3&0.0004&4&4&-0.028\\
3&8&1.285&4&3&0.030&2&5&-0.034\\
3&9&0.259&3&4&-0.031&4&5&-0.939\\
2&10&-0.939&3&5&-0.015&3&6&0.940\\
4&10&-0.016&3&7&1.328&2&7&-0.909\\
2&11&0.006&2&8&0.023&4&7&-0.016\\
4&11&-1.312&4&8&0.262&1&8&-0.194\\
1&12&0.940&2&9&0.010&1&9&0.888\\
&&&4&9&-1.302&\\
&&&1&10&0.016&\\
\end{tabular}
\end{ruledtabular}
\caption{
The 
electric-dipole
matrix
elements between 
the states of 
Fig.~\ref{fig:energyLevelsv0}
(in units of
$ea_0$)
for a magnetic field
of 
1000~gauss,
which defines the 
quantization axis.
The 
non-zero
$\Delta m_F=0$
($\pi$)
and 
$\Delta m_F=\pm1$
($\sigma^\pm$)
matrix elements are 
shown.
The similar 
matrix elements 
connecting to the 
$v=1$
states
are 
smaller by a factor 
of 
$\sqrt{0.035/0.964}$.
}
\label{tab:matrixEls}
\end{table}


\bibliography{polarbeam}

\end{document}